\documentstyle[12pt, epsf]{article}

\begin{document}

\font\mybbb=msbm10 at 8pt
\font\mybb=msbm10 at 12pt
\def\bbb#1{\hbox{\mybbb#1}}
\def\bb#1{\hbox{\mybb#1}}
\def\Z {\bb{Z}}
\def\pR{\bbb{R}}
\def\R {\bb{R}}
\def\C {\bb{C}}
\def\pC{\bbb{C}}
\def\H {\bb{H}}
\newcommand {\bm}[1]{\mbox{\boldmath $#1$}}
\newcommand {\al}{\alpha}
\newcommand {\del}{\delta}
\newcommand {\ga}{\gamma}
\newcommand {\eps}{\epsilon}
\newcommand {\la}{\lambda}
\newcommand {\om}{\omega}
\newcommand {\si}{\sigma}
\newcommand {\de}{\partial}
\newcommand {\lan}{\langle}
\newcommand {\ran}{\rangle}
\newcommand {\ra}{\rightarrow}
\newcommand {\be}{\begin{equation}}
\newcommand {\ee}{\end{equation}}
\newcommand {\bea}{\begin{eqnarray}}
\newcommand {\eea}{\end{eqnarray}}
\newcommand {\nona}{\nonumber}
\newcommand {\T}{\mathcal T}
\newcommand {\fa}{\forall\,}
\newcommand {\ex}{\exists\,}
\newcommand {\exi}{\exists !\,}
\newcommand {\pp}{\,\,\,}
\newcommand {\p}{\,\,}
\newcommand {\map}{\mapsto}
\newcommand {\cd}{\cdot}
\newcommand {\ba}{\begin{array}}
\newcommand {\ea}{\end{array}}
\newcommand {\per}{\times}
\newcommand {\ten}{\otimes}
\newcommand {\wed}{\wedge}
\newcommand {\E}{\mathcal E}
\newcommand {\da}{\dagger}
\newcommand {\A}{\mathcal A}
\newcommand {\cs}[1]{#1^{\dagger}}
\newcommand {\nome}{\label}
\newcommand {\bsi}{\bar{\si}}
\newcommand {\ama}{\left (\ba}
\newcommand {\cma}{\ea\right )}
\newcommand {\ie}{{\it i.e.}}
\newcommand {\en}{\mbox{End}}
\newcommand {\ho}{\mbox{Hom}}
\newcommand {\Ga}{\Gamma}
\newcommand {\mat}[1]{{\cal{#1}}}
\newcommand {\bib}{\bibitem}
\newcommand {\bz}{\bar{z}}
\newcommand {\bde}{\bar{\de}}
\newcommand {\na}{\nabla}
\newcommand {\D}{D^{\Gamma}}
\newcommand {\x}{d^{\mat{R}}\xi}
\def\id {\bm{1}}

\baselineskip =20pt
\thispagestyle{empty}
\vskip 0.5cm
\centerline{\huge \bf Gauge Zero-Modes on} 
\vskip 0.5cm
\centerline{\huge \bf ALE Manifolds} 
\vspace{1cm}
%%%%%%%%%%%%%%%%%%%%%%%%  AUTORI  %%%%%%%%%%%%%%%%%%%%%%%%%%%%%%
\centerline{\sc Cristiano Carpi}
\vskip 0.1cm
\centerline{\sl Dipartimento di Fisica, 
Universit\`a di Roma ``Tor Vergata",}  
\centerline{\sl Via della Ricerca Scientifica, 00133 Roma, ITALY}
\vskip 1cm
\centerline{\sc Francesco Fucito}
\vskip 0.1cm
\centerline{\sl INFN, sez. di Roma 2}  
\centerline{\sl Via della Ricerca Scientifica, 00133 Roma, ITALY}
\vskip 1cm
\centerline{\bf ABSTRACT}
{In this paper we find the general (i.e. valid for arbitrary values of the 
winding number) form of the gauge zero-modes, in the
adjoint representation, for theories living on manifolds of the ALE type.}

\pagenumbering{arabic}

\section{Introduction}
In the past few years there has been a considerable progress in the understanding
of non-perturbative effects in supersymmetric (SUSY) gauge field theories.
In the case of the theory with global $N=2$ SUSY, using a certain 
number of educated
guesses, all the non-perturbative contributions to the holomorphic part 
of the action have been calculated \cite{seiwitt}. 
Moreover, the results of \cite{seiwitt} have been generalized to a certain 
number of curved manifolds, to compute topological invariants of the Donaldson
type \cite{mw}.
A part of the computation in \cite{seiwitt} has been
checked by comparing with the results obtained by a saddle point
approximation of the functional integral around a self-dual solution 
(with winding numbers one and two) of the equations of motion of the theory
\cite{inst}. As of today, no checks have been performed on the results
in \cite{mw}.
While guess-work can be very powerful in certain occasions, the advantage
of a direct computation of the functional integral lies in the ease with which
it can be generalized to different situations. For example the breaking of SUSY
in supergravity theories by non-perturbative effects, leads to an explanation
of the generation of mass hierarchies, one of the most outstanding problems
in today's high-energy theoretical physics. The signature of these
non-perturbative effects is the formation of fermionic or bosonic condensates
which can be computed by the saddle point expansion we discussed before.
While the calculations in \cite{inst} were performed in flat space, in the
case of supergravity we need a generalization to curved
manifolds. These manifolds have to obey certain requirements if we want the 
classical supergravity theory to be a low-energy description of a 
heterotic string theory (or, that is the same, if we want to satisfy, the low-energy
equations of motion of the heterotic string).
The latter, in its turn, can act as an ultraviolet cut-off of the otherwise 
non-renormalizable supergravity. It turns out that ALE manifolds 
and self-dual gauge connections (of winding numbers bigger than one)
satisfy the necessary requirements \cite{BFMR}.
Some preliminary computations of the above mentioned
condensates, were performed in \cite{BFMR1}. In doing the actual computations,
an ingredient one can not do without is the explicit form of the zero-modes of
the gauge fields in the adjoint representation: deducing this expression
is the subject of this paper.
The final result will be valid for arbitrary winding numbers. 
Gauge connections of arbitrary winding numbers,
were first constructed in \cite{aty}, while ALE spaces (or gravitational instantons
as they are also known) of arbitrary winding numbers were built in \cite{Kr}.
Finally gauge connections of arbitrary winding numbers on ALE manifolds were
described in \cite{KN}. A short review of some of these results is given in
the first two section of this work. In the middle sections of the work we find 
the form of the zero-modes in a way that strictly resembles \cite{C,CG}.
In the last section we check the general form in a particular case already studied
in \cite{BFMR2}.

\section{Review of Kronheimer Construction of ALE Spaces}

Before facing the Kronheimer-Nakajima construction of all ADHM instantons on
ALE surfaces, we need to review the fundamental points of the Kronheimer
construction of ALE spaces \cite{Kr}. From the mathematical point of 
view, ALE manifolds are obtained exploiting a procedure called 
{\it hyper-K\"{a}hler quotient}. This procedure is a little involved and calls
for some explanations. 

The starting point is the set
\be\nome{Y} Y\equiv (\H^*\ten_{\pR}\mbox{End}(R))^{\Gamma}.\ee
$\mbox{End}(R)$ stands for the adjoint endomorphisms of 
the linear space $R$ of the
regular (adjoint) representation of the discrete group $\Gamma\subset SU(2)$.
$\H^*$ stands for the dual of the quaternion space $\H$. The action of
$\Gamma$ on $\H^*$ is induced by the usual action of $Sp(1)\sim SU(2)$ on $\H$.
The superscript $\Gamma$ in (\ref{Y}) means that we must choose $Y$ as the 
$\Gamma$-invariant subset of $(\H^*\ten_{\pR}\mbox{End}(R))$. 

To be explicit, $(\H^*\ten_{\pR}\mbox{End}(R))$ is 
the set composed by the matrices 
\be \nome{Y3}y=y^k\bsi_k,\pp k=1,2,3,4,\ee
where the $\bsi_i$'s are, respectively, the standard $2\times 2$ matrices 
$\id$, $-i\si_3^P$, $-i\si_2^P$, $-i\si_1^P$, and the $y^k$'s are
$|\Gamma|\times|\Gamma|$ adjoint matrices ($|\Gamma|$ is the dimension of
$\Gamma$) 
\be\nome{Y2}y=\ama{cc}y^1-iy^2 & -y^3-iy^4\\
                      y^3-iy^4 & y^1+iy^2\cma=
              \ama{cc}\al & -\cs{\beta}\\
                      \beta & \cs{\al}\cma,\ee
(\ref{Y2}) is the isomorphism 
\be\nome{iso}(\H^*\ten_{\pR}\mbox{End}(R))\sim\ho
(S^+\ten R,Q\ten R)_{\pR},\ee                  
where $S^+$ is isomorphic to $\C^2$ (in physicist's language it is the space acted 
upon by right-handed spinors)
and $Q$ is the linear space of the 
fundamental representation of $SU(2)$. Given a
$$\ga=\ama{cc}u & v\\-\bar{v} & \bar{u}\cma\in\Gamma\subset SU(2),$$ 
we can constrain $\al$ and
$\beta$ imposing
\be \nome{invarianza}R(\ga^{-1})\al R(\ga)=u\al+v\beta,\pp
    R(\ga^{-1})\beta R(\ga)=-\bar{v}\al+\bar{u}\beta,\ee
where $R(\ga)$ stands for the regular (adjoint) representation of $\ga$.

The set $Y$, equipped with the Euclidean metric
\be \nome{metrica}ds^2=\mbox{Tr}(dy\cs{dy}),\ee
is a flat manifold with hyper-K\"{a}hlerian structure in the sense of Calabi
\cite{Ca}. This means that we can define three covariantly costant 
endomorphisms of the tangent space $TY\equiv Y$, say $I,J,K$, respecting the 
quaternionic algebra 
\be I^2=J^2=K^2=-\id;\pp IJ=-JI=K.\ee
Looking at the expression (\ref{Y3}), it is easy to see that $I,J,K$ can be
chosen as 
\be I,J,K=\bsi_1,\bsi_2,\bsi_3\ten\id_R.\ee

The hyper-K\"{a}hler manifold $Y$ plays the role of an immersion space in the 
Kronheimer construction. To see this we have to note that the metric 
(\ref{metrica}) admits an isometry group $G\subset U(|\Gamma|)$ acting by the 
transformation law
\be \nome{transf}\ba{l}\al\map g\al\cs{g}\\
          \beta\map g\beta\cs{g}\ea,\ee
where $g\in G$ is any element of the unitary group $U(|\Gamma|)$ commuting 
with the action fo $\Gamma$ on $R$. Now, we are able to define the {\it moment maps} $\mu_i$, $i=1,2,3$, as the 
elements of $\mat{G}^*$ satisfying 
\be \ba{l}d(\mu_i\cd\lambda)={\om}_i(V_{\lambda})\ea,\ee
where $\lambda$ is any element of $\mat{G}$, $(\pp\cd\pp)$ is the internal 
product in $\mat{G}$, $V_{\lambda}$ is the Killing vector corresponding to 
$\lambda$ and ${\om_1},{\om_2},{\om_3}$ are the three closed 
K\"{a}hler 2-forms induced by the hyper-K\"{a}hler structure
\be\ba{l}{\om_1}=\frac{1}{2}\mbox{Tr}(dyI\cs{dy})\\ 
         {\om_2}=\frac{1}{2}\mbox{Tr}(dyJ\cs{dy})\\
         {\om_3}=\frac{1}{2}\mbox{Tr}(dyK\cs{dy})\ea.\ee

Explicitly, one can see that $\mu_1,\mu_2,\mu_3$ are the three $|\Gamma|\times
|\Gamma|$ traceless skew-adjoint matrices \cite{Kr}
\be\ba{l}\mu_1=\frac{1}{2}i([\al,\cs{\al}]+[\beta,\cs{\beta}])\\  
         \mu_2=\frac{1}{2}([\al,\beta]+[\cs{\al},\cs{\beta}])\\
         \mu_3=\frac{1}{2}i(-[\al,\beta]+[\cs{\al},\cs{\beta}])\ea.\ee
Choosing a suitable linear combination we can write
\be \ba{l}\mu_{\pC}=\frac{1}{2}[\al,\beta]\\
          \mu_{\pR}=\frac{1}{2}i([\al,\cs{\al}]+[\beta,\cs{\beta}])\ea.\ee
Equating $\mu_{\pC}$ and $\mu_{\pR}$, respectively, to $\zeta_{\pC}$ and 
$\zeta_{\pR}$, where the $\zeta$'s are parameters laying in the center of 
$\mat{G}^*$ (traceless matrices invariant under the action of $G$ given by 
(\ref{transf})), one obtains the ``level surfaces" $Y_{\zeta}$, 
\be \nome{Kr}\ba{l}\zeta_{\pC}=\frac{1}{2}[\al,\beta]\\
          \zeta_{\pR}=\frac{1}{2}i([\al,\cs{\al}]+[\beta,\cs{\beta}])\ea.\ee
The main result of \cite{Kr} is that varying $\zeta_{\pC}$ and $\zeta{\pR}$ 
and the group $\Gamma$ we can obtain all the hyper-K\"{a}hler 
four-manifolds with ALE structure as the manifold
\be\nome{ALE}X_{\zeta}=Y_{\zeta}/G.\ee
In particular, it is always possible to put $\zeta_{\pR}=0$ and, for every 
choosing of $\zeta_{\pC}\neq 0$, we obtain an ALE manifold resembling 
$\R^4/\Gamma$ at infinity.

More clearly, if we call $\xi$ the elements of $Y_{\zeta}$, the metric 
(\ref{metrica}) induces on $Y_{\zeta}$ the metric
\be\nome{metrica3} ds^2=\mbox{Tr}(d\xi d\cs{\xi}).\ee
Since the Kronheimer conditions (\ref{Kr}) are invariant under the action of 
$G$ given by (\ref{transf}), the metric (\ref{metrica3}) still possesses the 
isometry $G$. According to (\ref{ALE}) we can therefore obtain $X_{\zeta}$ 
by  gauging the $G$-invariance in (\ref{metrica3}). The net effect of this 
procedure turns out to be the substitution of $d\xi$ with  
\be\nome{taut}
d^{\mat{R}}\xi=(d+[A^{\mat{R}}_{\xi},\pp])\xi\ee in 
the metric of $Y_{\zeta}$, obtaining for $X_{\zeta}$ the metric 
\be\nome{Svolta}
ds^2_{X_{\zeta}}=\mbox{Tr}(d^{\mat{R}}\xi \cs{(d^{\mat{R}}\xi)}),\ee
where, matematically speaking, $d^{\mat{R}}$ is the covariant differentiation 
on the matrices $\xi$ saw as sections
\be \xi\in (\H^*\ten_{\pR}\mbox{End}(\mat{R}))^{\Gamma}\ee
of the so-called {\it tautological bundle} $\mat{R}$ obtained from the 
principal $G$-bundle $Y_{\zeta}$ as \cite{KN}
\be \mat{R}=Y_{\zeta}\times_G R.\ee

It comes from \cite{GN} that $\mat{R}$ has a natural decomposition
\be\nome{scomp}\mat{R}=\bigoplus_{i=0}^{r-1}\mat{R}_i\ten R_i,\ee
where the $R_i$'s are all the irreducible linear spaces of the 
representation of $\Gamma$ ($R_0$ is the trivial representation), and that the 
structure group of $\mat{R}$ turns out to be 
\be G=\bigotimes_{i\neq 0}^{r-1}U(|R_i|).\ee
In this way the connection $A^{\mat{R}}_{\xi}$ can be deduced according to the 
decomposition (\ref{scomp}) from the properties of the $A_i$ connections 
equipping the vector bundles $\mat{R}_i$. In particular \cite{GN} we have 
that $A_i$ possesses an antiself-dual curvature with finite action. 
As a connection in $\mat{G}$, $A^{\mat{R}}_{\xi}$ is a skew-Hermitian 
connection. This means that, with the help of (\ref{Y3}), we can 
write
\be \nome{Svolta2}ds^2_{X_{\zeta}}=2\sum_{k=1}^4\mbox{Tr}((d^{\mat{R}}\xi)^k 
(d^{\mat{R}}\xi)^k),\ee
where we used the identity
\be\nome{Fond}\bsi_k\si_i=i\bar{\eta}^a_{\pp ki}\si_a^P+\del_{ki}\id,\pp 
a=1,2,3.\ee    
The symbol $\bar{\eta}^a_{\pp ki}$ is the skew-symmetric, antiself-dual 't 
Hooft symbol.

(\ref{Svolta}) will be useful later.

\section{Kronheimer-Nakajima Construction}

Now we are able to face the Kronheimer-Nakajima (KN) construction of all 
ADHM instantons with topological index $k/|\Gamma|$ on ALE manifolds. 

The analogue of the $D$ matrix of the ADHM construction on flat spaces 
(see for example \cite{CG} and references therein) is 
\be\nome{D}D=(\mat{A}\ten\id_{R}-\id_{V}\ten\xi)\oplus (\Psi\ten\id_{R}),\ee
where
\be \mat{A}=\ama{cc}A & -\cs{B}\\
                     B & \cs{A}\cma,\pp \mat{A}\in(\H\ten_{\pR}\en(V)),\ee
and 
\be \Psi=\ama{cc}s & \cs{t}\cma,\pp s,\cs{t}\in\ho (V,W).\ee
We choose $V$ and $W$ as, respectively, $\C^k$ and $\C^n$ isomorphic
$\Gamma$-equivariant linear spaces, that is
\be V=\bigoplus_{i=0}^{r-1}R_i\ten V_i,\pp W=\bigoplus_{i=0}^{r-1}R_i\ten
W_i,\ee
where $V_i\sim\C^{v_i}$ and 
$W_i\sim\C^{w_i}$ are $\Gamma$-invariant spaces.

The matrix $D$ defined in eq. (\ref{D}), as an operator 
\be D: S^+\ten V\ten\mat{R}\ra 
(Q\ten V\ten\mat{R})\oplus(W\ten\mat{R}),\ee
is a $(2k+n)|\Gamma|\times 2k|\Gamma|$
matrix. We can obtain a 
$(2k+n)\times 2k$ matrix, as in the ADHM construction on flat spaces, 
simply reducing\footnote{In principle $\D$ is a 
$(2k+n)|\Gamma|\times 2k|\Gamma|$ matrix like $D$. The point is that 
starting from $\D$ it is always possible to cancel $|\Gamma|$ rows and 
columns without affecting all the KN construction.} 
$D$ to his $\Gamma$-invariant restriction $\D$ 
\be \nome{Dgamma}\D: S^+\ten (V\ten\mat{R})^{\Gamma}\ra 
(Q\ten V\ten\mat{R})^{\Gamma}\oplus(W\ten\mat{R})^{\Gamma}.\ee

The matrix $D^{\Gamma}$ must satisfy the ADHM conditions
\be\nome{cond2} \cs{(D^{\Gamma})}D^{\Gamma}=F^{-1}=f^{-1}\ten\id_{S^+}.\ee
(\ref{Kr}) and (\ref{cond2}) together give
\be\nome{cond}\ba{l}[A,B]+ts=\zeta_{\pR}\\

                    [A,\cs{A}]+[B,\cs{B}]-\cs{s}s+t\cs{t}=\zeta_{\pC}\ea,\ee
where, this time, $\zeta_{\pR}$ and $\zeta_{\pC}$ are 
\be \zeta=\bigoplus_{i=0}^{r-1}\zeta_i\id_{V_i},\ee
where $\sum_i\zeta_i=0$. 
Finally, the $SU(n)$ gauge bundle ${E}$ with instanton connection
$A_{\mu}^E$ and antiself-dual curvature $F^E_{\mu\nu}$, is given by    
\be\nome{E}E\equiv\mbox{Ker}\cs{(D^{\Gamma})}.\ee
The instanton connection $A_{\mu}^E$ is 
\be\nome{A_E} A_{\mu}^E=\cs{U}\na_{\mu}^{\mat{R}} U,\ee
where the matrix 
\be\nome{U2} U:E\ra(Q\ten V\ten\mat{R})^{\Gamma}\oplus(W\ten\mat{R})^{\Gamma}\ee  
is chosen in accordance with the conditions
\be\nome{U}\ba{l}\cs{(D^{\Gamma})}U=0\\

                 \cs{U}U=\id\ea\ee
and the derivation $\na_{\mu}^{\mat{R}}$ contains the Levi-Civita connection
and the connection 
$A_{\mu}^{\mat{R}}$ of the bundle $\mat{R}$ acting on $U$ according to 
(\ref{U2}). 

\section{Curvature in KN Construction}

To show the close analogy between KN and ADHM formalism on $\R^4$, we 
will verify the antiself-duality of the curvature given by the instanton 
connection (\ref{A_E}). Our proof will be basically the same given by 
\cite{C} in the case of the ADHM construction on $\R^4$.

The curvature $F_{\mu\nu}^E$ is given by the formula
\be \nome{F}F_{\mu\nu}^E=[\na_{\mu}^E,\na_{\nu}^E],\ee
where $\na_{\mu}^E$ is the covariant derivative with respect to the Levi-Civita
and $A_{\mu}^E$ connection given by (\ref{A_E}). From (\ref{A_E}) it follows that,
on any section $\phi$ of the bundle $E$, one has 
\be \nome{der}\na_{\mu}^E\phi=\cs{U}\na_{\mu}^{\mat{R}}(U\phi).\ee
Substituting (\ref{der}) into (\ref{F}) we find
\be \nome{bah}F_{\mu\nu}^E=\cs{U}\na_{[\mu}(U\cs{U}\na_{\nu]}U),\ee
where, for the sake of simplicity, we omitted the superscript $\mat{R}$ of the
covariant derivatives.

Expanding (\ref{bah}), we find 
\be F_{\mu\nu}^E=\cs{U}\na_{[\mu}P\na_{\nu]}U+\cs{U}F_{\mu\nu}^{\mat{R}}U,\ee
where we put $F_{\mu\nu}^{\mat{R}}=[\na_{\mu}, \na_{\nu}]$ and $P=\cs{U}U$.
Since 
\be 
P=\id-\D F\cs{(\D)},
\ee
we find 
\be \nome{gia'}F_{\mu\nu}^E=\cs{U}\na_{[\mu}\D
F\na_{\nu]}\cs{(\D)}U+\cs{U}F_{\mu\nu}^{\mat{R}}U,\ee
where we used the identity $\cs{(\D)}P=0$.

The covariant derivative $\na_{\mu}$ acts on $\D$ according to 
(\ref{Dgamma}). Looking at the definition of $D$ given in (\ref{D}), one 
can see that 
\be\nome{dxi}\na_{\mu}\D=-\cs{b}\na_{\mu}\xi,\ee
where, from now on, we abbreviate $\id_V\ten\xi$ with $\xi$ and we put $b$ 
for the projection to $(Q\ten V\ten\mat{R})^{\Gamma}$ in 
$(Q\ten V\ten\mat{R})^{\Gamma}\oplus(W\ten\mat{R})^{\Gamma}$ ($b$ is the 
analogue of the matrix which multiplies the coordinate $x\in\R^4$
in the ADHM construction in flat space \cite{CG})
(\ref{gia'}) then becomes
\be \nome{curv}F_{\mu\nu}^E=\cs{U}\cs{b}\na_{[\mu}\xi F\na_{\nu]}\cs{\xi}bU+
\cs{U}F_{\mu\nu}^{\mat{R}}U.\ee

From the properties of the bundle $\mat{R}$ described at the end of Sec. 
1 we see that $F_{\mu\nu}^{\mat{R}}$ is an antiself-dual quantity. The 
remaining part of  (\ref{curv}) is  more easily written in 
terms of differential 2-forms as
\be \nome{carino}\cs{U}\cs{b}\x \wed F\cs{\x}bU
=-\cs{U}\cs{b}(\x)^k\bsi_k \wed F(\x)^i\si_i bU.\ee

Looking at (\ref{Svolta2}), we see that, reducing $\x$ to a $2\times 2$ 
matrix, $(\x)^k$ 
can be normalized by a constant factor to a local 
orthonormal basis $e^k\in T^*X_{\zeta}$. The matrix $F$ satisfying the 
condition (\ref{cond2}) commutes with $\bsi_k$ and $\si_i$, so that, 
remembering (\ref{Fond}), it is easy to see that (\ref{carino}) also is an 
antiself-dual quantity.

\section{Bosonic Zero-Modes}

The bosonic zero-modes of a YM theory with antiself-dual curvature 
are determined \cite{Os} by  
\be\nome{ZM1}\na^E_{ad\p[\mu}Z_{\nu]}=-*\na^E_{ad\p[\mu}Z_{\nu]}\ee
and 
\be\nome{ZM2}\na^E_{ad\p\mu}Z^{\mu}=0,\ee
where, since $Z_{\mu}$ is in the adjoint representation of the
gauge group $SU(n)$, we have to use here $\na^E_{ad\p\mu}$ instead of $\na^E_{\mu}$. 
The symbol $*$ stands for the duality operator.

We choose for $Z_{\mu}$ the form 
\be\nome{sol1}Z_{\mu}=\cs{U}\cs{z}_{\mu}-z_{\mu}U,\ee
where 
\be\nome{sol2}z_{\mu}=-\cs{U}\na_{\mu}D^{\Gamma}F\cs{C}.\ee
$\cs{C}$ is a costant matrix,
\be \cs{C}:(Q\ten V\ten\mat{R})^{\Gamma}\oplus(W\ten\mat{R})^{\Gamma}\ra
S^+\ten (V\ten\mat{R})^{\Gamma},\ee
satisfying the condition 
\be \na_{\mu}\cs{C}=0.\ee
This is equivalent to say that  
\be \cs{C}=((\mat{B}\ten\Phi)\ten\id_R)^{\Gamma},\ee
with 
\be \mat{B}\in\ho_{\Gamma}(S^+\ten V, Q\ten V)_{\pR}\ee
and 
\be \Phi\in\ho_{\Gamma}(V,W).\ee

To demonstrate that $Z_{\mu}$ given by (\ref{sol1}) is a solution
of (\ref{ZM1}) and (\ref{ZM2}), we need the explicit expression of 
$\na^E_{ad\p\mu} Z_{\nu}$. Since, using (\ref{A_E}),
\be \na^E_{ad\p\mu} Z_{\nu}=\cs{U}\na_{\mu}(UZ_{\nu}\cs{U})U,\ee
we can write
\bea\nome{boh}
\na^E_{ad\p\mu} Z_{\nu}
&=&\cs{U}\na_{\mu}(P\cs{z_{\nu}}\cs{U}-Uz_{\nu}P)U\\
&=&\cs{U}\cs{b}\na_{\mu}\xi F\cs{(\D)}\cs{z_{\nu}}+
\cs{U}\na_{\mu}(\cs{z_{\nu}}\cs{U})U+\\
\nona &-&\cs{U}\na_{\mu}(Uz_{\nu})U-z_{\nu}\D F\na_{\mu}\cs{\xi}b U=\\
&=&K_{\mu\nu}+\cs{U}\cs{(\na^E_{\mu}z_{\nu})}-\na^E_{\mu}z_{\nu}U.
\eea

Now, to simplify the proof of (\ref{ZM1}) and (\ref{ZM2}) it is useful
to note that the quantity $z_{\mu}$ is a solution of 
\be \nome{ZMF1}
\na^E_{\mu}z_{\nu}=-*\na^E_{\mu}z_{\nu}
\ee
and 
\be \nome{ZMF2}
\na^E_{\mu}z^{\mu}=0.
\ee

(\ref{ZMF1}) is easily checked by writing explicitly the derivative
$\na^E_{\mu}z_{\nu}$ and proceeding as in Sec. 3. The proof of
(\ref{ZMF2}) is a little more involved. First of all we have to note that, 
setting $\cs{C}\equiv (\cs{C})^{\al}\chi_{\al}$, 
where the index $\al=1,2$ spans the $S^+$ space, it
is always possible to write
\be\nome{fico} z_{\mu}=z_{\mu}^{\al}\chi_{\al}.\ee
From (\ref{fico}) we can construct
the left-handed spinor
\be \bar{\lambda}_{\dot{\al}}=z_{\mu}^{\al}\si^{\mu}_{\pp\al\dot{\al}},\ee
which is a solution \cite{KN} of the Dirac equation   
\be
\widetilde{\na}^E_{\nu}(z_{\mu}^{\al}\si^{\mu}_{\pp\al\dot{\al}})
\bsi^{\nu\p\dot{\al}\beta}=0,
\ee
where $\widetilde{\na}^E_{\nu}$ is the covariant derivative with respect to the
Levi-Civita, spin and instanton $A^E_{\mu}$ connection.

Since on a self-dual background, like ALE manifolds, we have 
\be\widetilde{\na}^E_{\nu}(z_{\mu}^{\al}\si^{\mu}_{\pp\al\dot{\al}})
\bsi^{\nu\p\dot{\al}\beta}=
\na^E_{\nu}z_{\mu}^{\al}\si^{\mu}_{\pp\al\dot{\al}}\bsi^{\nu\p\dot{\al}\beta},\ee
we can write
\be \nome{svolta}\na^E_{\nu}z_{\mu}\si^{\mu}\bsi^{\nu}=
\na^E_{\nu}z_{\mu}(i\eta_k^{\mu\nu}\si^k+g^{\mu\nu})=0,\ee
where $\eta_k^{\mu\nu}$ is the 't Hooft self-dual symbol and $g^{\mu\nu}$ is
the metric.

From (\ref{ZMF1}) and (\ref{svolta}), it comes that 
\be \na^E_{\nu}z_{\mu}g^{\mu\nu}=0,\ee
which is (\ref{ZMF2}).

Using (\ref{ZMF1}) and (\ref{ZMF2}), (\ref{ZM1}) and (\ref{ZM2}) reduce
then to 
\be \nome{1}K_{[\mu\nu]}=-*K_{[\mu\nu]}\ee
and 
\be \nome{2}K_{\mu}^{\pp\mu}=0.\ee
From (\ref{boh}) it comes that
\be K_{[\mu\nu]}=\cs{U}F\cs{b}\na_{\mu}\xi(\cs{(\D)}
C+\cs{C}\D)\na_{\nu}\cs{\xi}b U-\mbox{h.c.}\ee
Comparing this expression with  (\ref{curv}), one sees that 
\be \nome{wow}\cs{(\D)}C+\cs{C}\D=G^{-1}=g^{-1}\ten\id_{S^+},\ee
is a sufficient condition to assure the antiself-duality of $K_{[\mu\nu]}$.

(\ref{wow})is identical to the $U(n)$ version of the condition 
found in \cite{Os} for the bosonic zero-modes on $\R^4$.

For what (\ref{2}) is concerned,, we see that 
\be K^{\mu}_{\pp\mu}=\cs{U}\cs{b}\na^{\mu}\xi(\cs{(\D)}C-\cs{C}\D)
\na_{\mu}\cs{\xi}bU.\ee
Remembering that $\x$ can be written as $e^k\bsi_k$ and that $e^k$ is a 
local vierbein basis in $T^*X_{\zeta}$, we obtain
\be K^{\mu}_{\pp\mu}=\cs{U}\cs{b}f\bsi^k(\cs{(\D)}C-\cs{C}\D)
\si_k bU.\ee
(\ref{wow}) means that $\cs{(\D)}C$ must be a matrix of the 
form $$\ama{cc}\ga & -\cs{\del}\\ \del & \cs{\ga}\cma.$$ Since for this kind 
of matrix we have
\be\sum_{k=1}^4\left (\bsi^k\ama{cc}\ga & -\cs{\del}\\ \del & \cs{\ga}\cma\si_k
\right )=2(\ga+\cs{\ga})\ten\id_{S^+},\ee
it is easy to verify that
\be K^{\mu}_{\pp\mu}=0.\ee

\section{Gauge Zero-modes for $k=1/2$ on the Eguchi-Hanson Manifold}

As an example of (\ref{sol1}) and (\ref{sol2}) we compute
the zero-modes of the self-dual gauge 
potential of topological index 
$k=1/2$ corresponding to the choice $\Gamma\equiv\Z_2$. 

As it is explained in \cite{BFMR}, in this case the ADHM-KN construction gives 
the simplest instanton connection possible on the Eguchi-Hanson (EH) 
manifold. 
As a first step, it is necessary to determine the expression of $\xi$ and, 
consequently the EH metric from the Kronheimer construction described in 
Sec. 1. The choice  
$\Gamma\equiv\Z_2$ means that $R$ is isomorphic to $\C^2$. In 
the decomposition (\ref{scomp}) only $R_0$ and $R_1$ survive. This means that 
$A^{\mat{R}}$ 
can be written as
\be\nome{franz}  A^{\mat{R}}=\ama{cc}0 & 0\\0 & A^{U(1)}\cma,\ee
where $A^{U(1)}$ is a suitable abelian connection with antiself-dual 
curvature. Acting on $\xi$, which is a section of $\H^*\ten\mbox{End}
(\mat{R})$, (\ref{franz}) becomes
\be A^{\mat{R}}_{\xi}=\ama{cc}A^{\mat{R}} & 0\\0 & A^{\mat{R}}\cma.\ee

Furthermore, one can see \cite{BFMR} that
\be \xi=\ama{cccc}0 & v^1 & 0 & -\la\bar{v}^2\\
                  \la v^1 & 0 & -\la{v}^2 & 0\\
                  0 & v^2 & 0 & \la\bar{v}^1\\
                  \la v^2 & 0 & \la{v}^1 & 0\cma,\ee
where $v^1,v^2\in\C$ and $\la=1+a^2/\sum_{i=1}^2|v^i|^2$, so that $\x$ 
(reduced to a $2\times 2$ matrix) can be written as 
\be \x=(d+A^{U(1)})\ama{cc}v^1 & -\la\bar{v}^2\\
                           v^2 & \la\bar{v}^1\cma.\ee
The vierbein basis $e^k$ defined at the end of Sec. 3 is then
\be \nome{base}e^k=(d+A^{U(1)})\xi^k,\ee
where
\be\ba{cc}\xi^1=\frac{1}{2}(v^1+\la\bar{v}^1) & \xi^3=\frac{1}{2}(v^2+\la\bar{v}^2)\\
\xi^2=\frac{i}{2}(v^1-\la\bar{v}^1) & \xi^4=\frac{i}{2}(v^2-\la\bar{v}^2)\ea.\ee

Switching to coordinates 
\be\ba{l}v^1=\sqrt{\frac{r^2-a^2}{2}}\cos(\frac{\theta}{2})
e^{i\left (\frac{\psi+\phi}{2}\right )}\\
         v^2=\sqrt{\frac{r^2-a^2}{2}}\sin(\frac{\theta}{2})
e^{i\left (\frac{\psi-\phi}{2}\right )}\ea\ee
and choosing $A^{U(1)}$ (in the same coordinates) as the monopole 
potential \cite{EH} 
\be A^{U(1)}=i\frac{a^2}{r^2}\frac{(d\psi+\cos\theta d\phi)}{2}=
-i\frac{a^2}{r^2}\si_z,\ee
we find that $ds^2_{X_{\zeta}}=\sum_k(e^k)^2$ gives the EH metric.

Incidentally, we note that in the limit $a^2\ra 0$ 
($X_{\zeta}\ra\R^4/\Z_2$) the basis $e^k\equiv(\x)^k$ reduces 
to the canonical basis of differential 1-forms in $\R^4$, $dx^1,...,dx^4$, 
with $x^1,...,x^4\equiv\xi^1,...,\xi^4\in\R^4$.

Knowing the expression of the matrix $\xi$ we are able to find 
$D$ and, consequently, $\D$. One can see \cite{BFMR} that 
\be \nome{Dbella}\D=\ama{cc}v^1 & -\la\bar{v}^2\\
               v^2 & \la\bar{v}^1\\
               s^1 & -\mu\bar{s}^2\\
               s^2 & \mu\bar{s}^1\cma,\ee
with $s^1,s^2\in\C$ and $\mu=1-{a^2}/({\sum_{i=1}^2|s^i|^2})$.
From (\ref{Dbella}), solving (\ref{wow}), we can find the
expression of the matrix $C$ 
determining the bosonic zero-modes of the instanton potential. 

It is easy to see that $C$ can be chosen as
\be C=\frac{1}{2}\ama{cc}0 & 0\\0 & 0\\s^1 & {\bar{s}^2}/{\mu}\\
                         s^2 & -{\bar{s}^1}/{\mu}\cma,\ee
so that
\be\cs{(\D)}C+\cs{C}\D=\sum_i|s^i|^2\ten\id_{S^+}.\ee
From (\ref{Dbella}) we can calculate $U$, which turns out to be
\be U=\frac{|s|}{|v|\sqrt{|v|^2+|s|^2}}\ama{cc}v^1 & \mu\bar{v}^2\\
                                       v^2 & -\mu\bar{v}^1\\
                        -\frac{|v|^2}{|s|^2}s^1 & -\la\frac{|v|^2}{|s|^2}\bar{s}^2\\
                        -\frac{|v|^2}{|s|^2}s^2 & \la\frac{|v|^2}{|s|^2}\bar{s}^1\cma
,\ee
where we put $|s|^2=\sum_i|s^i|^2$, $|v|^2=\sum_i|v^i|^2$, and $F$.
As a consequence 
\be F=(|v|^2+|s|^2)\ten\id_{S^+}.\ee

Putting all the pieces of our construction together as in (\ref{sol1}) 
and (\ref{sol2}), we find for $Z=Z_{\mu}dx^{\mu}$ in EH coordinates the 
expression
\be \nome{esempio}Z=2i\frac{t^2+a^2}{\sqrt{t^4-a^4}}\ama{cc}f^3\si_z & f^1(\si_x-i\si_y)\\   
                                     f^1(\si_x+i\si_y) & -f^3\si_z\cma,\ee
where $t^2=2|s|^2+a^2$ and
\be f^1=\frac{t^2r^2+a^4}{(r^2+t^2)^2},\pp 
f^3=\frac{(r^4-a^4)\sqrt{t^4-a^4}}{r^2(r^2+t^2)^2}.\ee 

(\ref{esempio}) gives the bosonic zero-modes, already found in 
\cite{BFMR} with different methods.

\enddocument